\documentclass[pdftex,twocolumn,epjc3]{svjour3}          

\RequirePackage[T1]{fontenc}

\smartqed  

\RequirePackage{graphicx}
\RequirePackage{mathptmx}      
\RequirePackage{flushend}
\RequirePackage[numbers,sort&compress]{natbib}
\RequirePackage[colorlinks,citecolor=blue,urlcolor=blue,linkcolor=blue]{hyperref}
\RequirePackage{paralist}
\RequirePackage{calrsfs}
\DeclareMathAlphabet{\pazocal}{OMS}{zplm}{m}{n}
\RequirePackage{multirow}
\RequirePackage{hhline}
\RequirePackage{amsmath}
\RequirePackage{amssymb}

\journalname{Eur. Phys. J. C}

\begin{document}

\title{Electromagnetic extension of Buchdahl bound in $f(R,T)$ gravity
}


\author{Soumik Bhattacharya\thanksref{e1,addr1}
        \and
        Ranjan Sharma\thanksref{e2,addr1} 
        \and
        Sunil D. Maharaj\thanksref{e3,addr2}
}

\thankstext{e1}{e-mail:soumik.astrophysics@gmail.com}
\thankstext{e2}{e-mail:rsharma@associates.iucaa.in}
\thankstext{e3}{e-mail:maharaj@ukzn.ac.za}

\institute{IUCAA Centre for Astronomy Research and Development (ICARD), Department of Physics, Cooch Behar Panchanan Barma University, Cooch Behar, 736101, India.\label{addr1}
          \and
          Astrophysics Research Centre, School of Mathematics, Statistics and Computer Science, University of KwaZulu-Natal, Private Bag 54001, Durban 4000, South Africa\label{addr2}
}

\date{Received: date / Accepted: date}

\maketitle

\begin{abstract}
We develop a static charged stellar model in $f(R,T)$ gravity where the modification is assumed to be linear in $T$ which is the trace of the energy momentum tensor. The exterior spacetime of the charged object is described by the Reissner-Nordstr\"om metric. The interior solution is obtained by invoking the Buchdahl-Vaidya-Tikekar ansatz, for the metric potential $g_{rr}$, which has a clear geometric interpretation. A detailed physical analysis of the model clearly shows distinct physical features of the resulting stellar configuration under such a modification. We find the maximum compactness bound for such a class of compact stars which is a generalization of the Buchdahl bound for a charged sphere described in $f(R,T)$ gravity. Our result shows physical behaviour that is distinct from general relativity.
\end{abstract}

\section{\label{sec1}Introduction}

One of the main pillars of modern physics for understanding the present universe is the general theory of relativity (GTR), discovered by Einstein in $1915$, which was first experimentally verified by Eddington in $1920$, and later by various tests in the solar system. Note that GTR was first modified very soon after its discovery. In $1919$, Weyl \cite{Weyl1919101}  introduced higher order invariants in the Einstein-Hilbert (EH) action unifying electromagnetism (EM) and gravity. Later, Kaluza and Klein \cite{Kaluza1921966,Klein1926895} investigated the higher dimensional effects on EM. Although the complexity of EH action had no apparent experimental motivations during that period, around the $1960$'s, many investigators found virtue in such an approach. Based on Kaluza's unitary field theory, Brans \textit{et al} \cite{Brans1961925} introduced a scalar-tensor theory from the observation of the solar oblateness and the precession of Mercury's orbit \cite{Dicke1967313} which was later taken up by  Bergmann \cite{Bergmann196825}.

The recent shreds of evidence emerging from astrophysics and observational cosmology suggest that the cosmic acceleration of the universe may have occurred in two phases. Preceding radiation domination, the inflationary phase \cite{Starobinsky198099} took place, which not only solves the flatness problem \cite{Guth1981347} in standard cosmology, but also justifies the nearly flat spectrum of temperature anisotropies observed in cosmic microwave background (CMB) \cite{Smooth1992L1}. The second phase is the matter-dominated present universe, i.e. the late-time acceleration supposedly originated from dark energy \cite{Huterer1999081301}. A supernova search team has experimentally verified the late-time expansion of the universe \cite{Riess19981009} based on the observational data of $10$ new high-redshift Type Ia supernovae as well as through rigorous and detailed experiments \cite{Perlmutter20121127} considering the brightness of supernovae as an indicator. Subsequently, the existence of a small but non-zero cosmological constant \cite{Perlmutter1999565} has been justified by the current mass-energy density of the universe. Moreover, when the information coming from anisotropies in CMB is combined with measurements of the light chemical element abundances on a cosmological scale, one concludes that about one-fifth of our universe is composed of non-luminous and non-baryonic material called dark matter \cite{Carroll20011,Peebles2003559,Riess2004665,Eisenstein2005560,Hooper2009709}. Hence, despite remarkable success in predicting many tests of gravitational phenomena including the most recent discovery of gravitational waves, GTR faces many challenges on several fronts - both on small and large scales.  

While scientific curiosity on the theoretical front provides ample motivation to contemplate modifying Einstein's gravity, the above observational evidence strongly justifies such exercises. GTR can be modified by adopting the Einstein-Hilbert action containing the Lagrangian density  $\sqrt{-g}R$, $R$ being the curvature scalar. The initial de-Sitter state of the universe is explained by adding a term proportional to $\sqrt{-g}R^{m}$ ($m > 0$) to the action, known as Starobinsky inflation \cite{Starobinsky198099}. However, if $m < 0$, then the acceleration of the universe originates from gravitation as shown by Carroll \textit{et al} \cite{Carroll2004043528}. The $\Lambda$~cold dark matter ($\Lambda$CDM) model, based on inflationary theory, explains the acceleration of the universe where one adds the cosmological constant $\Lambda$ to the EH action \cite{Nojiri2007238}, which fits well with several observational data \cite{Komatsu2009330}, and thereby also offers the possibility for time varying equation of state of dark energy \cite{Zhao2010043518}. However, the $\Lambda$CDM model is burdened with magnitude problems. A more radical alternative approach is the modified $f(R)$ gravity, which allows a generalization of the EH action to interpret some of the basic characteristics of the higher order curvature gravity. Before experimental observation, treating both the metric and the affine parameters as independent variables, Palatini \cite{Palatini1919203} formulated a different perspective of $f(R)$ gravity leading to second order field equations free from the instability associated with negative signs of the second order partial derivative of $R$ in the functional $f(R)$. However, Palatini's formulation turned out to be inconsistent with the late-time cosmic acceleration. Consequently, the formulation demanded a different approach to $f(R)$ theory. To obtain a divergence-free Einstein equation, Lanczos \cite{Lanczos1938842} proposed a specific combination of curvature-squared terms offering a modified theory named Einstein-Gauss-Bonnet (EGB) gravitational theory. Different cosmological models have been developed based on EGB gravity  \cite{Tangherlini1963636}. In higher dimensions, the possible existence of a black hole was shown by including a $4$D Gauss-Bonnet (GB) term to the EH action \cite{Boulware19852656}. Investigators have also reconciled the early-time inflation, and late-time acceleration of the universe, in EGB gravity \cite{Nojiri2007238,Nojiri20051,Cognola2007086002,Brevik2007817}. Recently, EGB gravity has also received a widespread application in astrophysics. Many researchers have interpreted the physical quantities in five-dimensional framework of EGB gravity assuming different kinds of interior geometries and fluid distributions \cite{Dadhich2010104026,Maharaj2015084049,Brassel2020971,Tangphati2021136423,Soumik20232350018}. In addition to EGB gravity, $f(R)$ theory \cite{Capozziello2006135,Nojiri2007115,Martins20071103,Boehmer2008024}(and references therein) explains how the cosmological constant can be bypassed geometrically \cite{Barrow19832757} by adding higher order curvature scalar to the action. 

To explain the current expansion of the universe as well as the dark energy scenario, the action in $f(R)$ is further extended by coupling non-minimally the matter field to the geometry (viz. the Ricci scalar $R$) \cite{Bartolami2007104016}, which leads to $f(R,T)$ theory, $T$ being the trace of stress-energy tensor. Thus, the associated continuity equation takes a different form which justifies the energy exchange between the matter and geometry beyond the curved spaces. This coupled matter part generates an additional force term orthogonal to the four velocities of a massive object connoting a non-geodesic nature of motion. The non-vanishing divergence of the stress tensor violates the equivalence principle which, however, can be controlled by the corresponding coupling parameter. Considering matter as a perturbation to a locally flat spacetime, it can be shown that the extension mentioned above satisfies the equivalence principle \cite{Sotiriou2008205002}. Following this, Harko \textit{et al} \cite{Harko2010044021} developed a model for stellar configuration filled with a perfect fluid. Various classes of solutions in $f(R, T)$ gravity have been discussed by Harko \textit{et al} \cite{Harko2011024020}. Note that the action in $f(R,T)$ theory is not Lorentz invariant and does not provide a frame-independent gravitational theory. The theory can only be formulated properly in a strong gravity regime by considering a linear functional form of $f(R,T)$ \cite{Cemsinan2011} for a spherically symmetric stellar configuration. With the linear functional form of $f(R,T)$, the anisotropic behaviour of a collapsing object \cite{Noureen201562,Zubair2017169}, effect of charge \cite{Lemos201576,Arbanil2018104045,Pretel2022115103}, and the stability criterion \cite{Deb2018084026,Deb20195652,Maurya2019044014,Jose2014028501} have been studied. 

In the recent past, many investigators have analyzed the gross physical properties of a stellar configuration by considering different stellar solutions in $f(R, T)$ gravity, viz. Tolman-IV solution \cite{Bhar20222201}, embedded class I solution with Karmakar condition \cite{Asghar2023427}, Krori-Barua type compact stellar solution \cite{Shamir20222200134}, and solutions obtained by utilizing Buchdahl's ansatz \cite{Maurya2020100438,Kumar2021100880,Bhar2023101990}. The effect of the parameter that couples the matter contribution to geometry on the interior structure of a compact object (e.g. neutron star) has been investigated by Pappas \textit{et al} \cite{Pappas2022124014} for the Tolman-VII solution in linear $f(R, T)$ gravity. The investigation also provided an analysis of the logical extension for the uniform density configuration. The analyses in ref.~ \cite{Pappas2022124014} also facilitated an examination of the upper bound on the mass to radius ratio in an appropriate parametric regime. Motivated by these developments, we are interested in obtaining an upper bound on the compactness of a charged sphere in $f(R, T)$ gravity analogous to the Buchdahl bound in GTR. Note that a charged generalization of the Buchdahl bound in Einstein's gravity was provided by Sharma \textit{et al} \cite{Sharma202179}. In this paper, we attempt to analyze the functional dependency of the coupling parameter on the compactness bound for a charged sphere filled with isotropic fluid by obtaining a solution in $f(R,T)$ gravity which is linear in $T$. The solution is obtained by introducing the Buchdahl-Vaidya-Tikekar \textit{ansatz} \cite{Buchdahl19591027,Vaidya1982325} as one of the metric potentials of the spherically symmetric static distribution. By invoking a particular coordinate transformation, we solve the system of equations and fix the constants of the solution by matching the interior solution to the exterior Reissner-Nordstr\"om (RN) metric across the boundary which facilitates its physical analysis. 

The structure of this paper is as follows: In Sec.~(\ref{sec2}), the effective field equations are obtained for a charged object with perfect fluid distribution enclosed in a sphere in $f(R, T)$ gravity theory. Sec.~(\ref{sec3}) deals with the technique for generating solutions of the field equations. The section is subdivided into two classes: (i) uncharged VT model, which reduces to the Schwarzschild incompressible interior solution for a particular choice of the model parameters; and (ii) charged VT model. The unknown constants are fixed in Sec.~(\ref{sec4}) by matching the interior solution to the exterior RN metric across the boundary. Gross physical properties are discussed in Sec.~(\ref{sec5}). Sec.~(\ref{sec6}) is devoted to the analysis of the modification on the compactness bound of the charged sphere. Some concluding remarks are made in Sec.~(\ref{sec7}). 

\section{\label{sec2}Electromagnetic formulation of field equations}

To study the stellar configurations of a charged compact star, Einstein's gravity is modified in conjunction with the trace of the stress-energy tensor $T$ to the existing Ricci scalar term $R$. The subsequent action has the form
\begin{equation}
\hspace{1.5cm}S = \int{\Big[\frac{f(R,T)}{16\pi}+\pazocal{L}_m +\pazocal{L}_e\Big]\sqrt{-g}d^4 x}, \label{act}
\end{equation}
$\pazocal{L}_m$ being the Lagrangian matter density and $g$ is the determinant of the metric tensor $g_{ab}$ \cite{Pretel2022115103}. \footnote{All the indices considered in this paper runs from $0$ to $3$.} In (\ref{act}), $\pazocal{L}_e$ represents the electromagnetic field Lagrangian density given by
\begin{equation}
\hspace{1.5cm}\pazocal{L}_e = j^{a}A_{a} - \frac{1}{16\pi}F_{ab}F_{cd}g^{ac}g^{bd}, \label{efl}
\end{equation} 
where the four-current density is given by $j^{a} = \sigma u^{a}$, with $\sigma$ being the electric charge density and $u^{a}$ is the four-velocity of the fluid satisfying the relations $u_{a}u^{b}=1$ and $u^{a}\nabla_{b}u_{a}=0$.  $F_{ab}=\nabla_{a}A_{b}-\nabla_{b}A_{a}$ is the electromagnetic field strength tensor where $A_{a}$ represents the electromagnetic four potential and $\nabla_{a}$ denotes the covariant derivative associated with the Levi-Civita connection of metric tensor $g_{ab}$. 

Thus, the total stress-tensor is a sum of two terms, namely the matter part $(\pazocal{M}_{ab})$ and the electromagnetic part $(\pazocal{E}_{ab})$, i.e.
\begin{equation}
\hspace{3cm}T_{ab}=\pazocal{M}_{ab} + \pazocal{E}_{ab}. \label{tft}
\end{equation}
In this paper, we consider a perfect fluid distribution, which implies $$\pazocal{M}_{ab}=(p+\rho)u_{a}u_{b} + pg_{ab},$$ where $\rho$ is the matter density and $p$ is the isotropic pressure, and the electromagnetic field tensor takes the form $$\pazocal{E}_{ab}=\frac{1}{4\pi}(F_{ac}F^{c}_{b}-\frac{1}{4}g_{ab}F_{cd}F^{cd}).$$
The Lagrangian matter density is related to the energy ~- momentum tensor as
\begin{equation}
\hspace{1.5cm}\pazocal{M}_{ab} = - \frac{2}{\sqrt{-g}} \frac{\delta(\sqrt{-g} \pazocal{L}_m)}{\delta g^{ab}}, \label{emt1}
\end{equation} 
where it is assumed that $\pazocal{L}_m$ depends only on the metric and not on its derivatives. The above equation on contraction yields 
\begin{equation}
\hspace{1.5cm}\pazocal{M}_{ab} = g_{ab} \pazocal{L}_m - 2\frac{ \partial \pazocal{L}_m}{\partial g^{ab}}. \label{memt}
\end{equation}
On variation of action (\ref{act}), with respect to the metric components $g^{ab}$, yields the following relationship
\begin{eqnarray}
\delta S &=& \frac{1}{16 \pi} \int \Bigl[(R_{ab}+g_{ab}\Box-\nabla_{a}\nabla_{b})f_{R}(R,T)\delta g^{ab} \Bigr.\nonumber\\ 
&& \Bigl. +f_{T}(R,T)\frac{\delta(g^{cd}\pazocal{M}_{cd})}{\delta g^{ab}}\delta g^{ab}
-\frac{1}{2} g_{ab}f(R,T)\delta g^{ab} \Bigr.\nonumber\\
&& \Bigl. +16\pi\frac{1}{\sqrt{-g}}\frac{\delta(\sqrt{-g}L_{m})}{\delta g^{ab}}\Bigr]\sqrt{-g}d^{4}x, \label{act1}	
\end{eqnarray}
where $f_{R}(R,T)=\frac{\partial f(R,T)}{\partial R}$, $f_{T}(R,T)=\frac{\partial f(R,T)}{\partial T}$ and $\Box \equiv$\\
$\frac{1}{\sqrt{-g}}\partial_{a}(\sqrt{-g}g^{ab}\partial_{b})$. As the trace of the electromagnetic energy-momentum tensor vanishes, the variation of the trace of the stress tensor with respect to the metric tensor is given by
\begin{equation} 
\hspace{1.5cm}\frac{\delta (g^{cd}\pazocal{M}_{cd})}{\delta g^{ab}}=\pazocal{M}_{ab}+\Theta_{ab}, \label{emt2}
\end{equation}
where $\Theta_{ab}\equiv g^{cd}\frac{\delta\pazocal{M}_{cd}}{\delta g^{ab}}$. On simplification of Eq.~(\ref{act1}), and using Eq.~(\ref{emt2}), one obtains the modified form of Einstein's field equations in $f(R,T)$ theory as
\begin{equation}
\Big(R_{ab}+g_{ab}\Box-\nabla_{a} \nabla_{b}\Big)f_{R}(R,T)-\frac{1}{2}f(R,T)g_{ab} 
= T_{ab}^{eff}, \label{mefe}
\end{equation}
where $T_{ab}^{eff}\equiv8\pi T_{ab}-f_{T}(R,T)(\pazocal{M}_{ab}+\Theta_{ab})$.

On account of all the terms contained in the conservation law \cite{Jose2014028501}, performing the covariant derivative of the right hand side of Eq.~(\ref{mefe}), and re-arranging, we obtain
\begin{eqnarray}
\nabla^{a} \pazocal{M}_{ab} &=& \frac{f_{T}(R,T)}{8\pi-f_{T}(R,T)}\Bigl[\left(\pazocal{M}_{ab}+
\Theta_{ab}\right)\nabla^{a}lnf_{T}(R,T) \Bigr.\nonumber\\ 
&& \Bigl.+\nabla^{a}\Theta_{ab}-\frac{1}{2}g_{ab}\nabla^{a}T-\frac{8\pi}{f_{T}(R,T)}\nabla^{a}\pazocal{E}_{ab}\Bigr]. \label{cdemt}
\end{eqnarray} 
The right hand side of the above equation vanishes if $f_{T}(R,T)=0$, i.e. $f(R,T)$ is simply a function of the Ricci scalar $R$, and hence one can easily retrieve the conservation law in Einstein gravity. The non-vanishing right hand term of Eq.~(\ref{cdemt}) was shown to play a crucial role in explaining the gravitational effects on the solar system beyond GR \cite{Harko2011024020}. 

Motivated by the growing interest of $f(R,T)$ gravity in the high gravity regime, we would like to develop and study the distinctive features of a charged fluid sphere in $f(R,T)$ gravity. Even though different functional forms of $f(R,T)$ have been explored in the past \cite{Harko2011024020}, Cemsinan \textit{et al} \cite{Cemsinan2011} showed that the only possible and acceptable form of $f(R,T)$ is a linear functional form of $T$ which can provide a plausible relativistic compact star model in $f(R,T)$ gravity. Hence, we write the modification in the form $f(R,T) =R+2\chi T $, where $ \chi $ is a dimensionless coupling parameter. Further, by choosing the matter Lagrangian density as $\pazocal{L}_m = p$ \cite{Pappas2022124014} (opposite signature was taken in ref.~\cite{Moraes2016005}), we have $\Theta_{ab}=-2 \pazocal{M}_{ab}$ \\
$+ p g_{ab}$ and it's trace provides $\Theta = -2 \pazocal{M}+4p$. Utilizing (\ref{mefe}), we eventually obtain  
\begin{equation}
\hspace{1cm}G_{ab}= 8\pi T_{ab}+\chi \pazocal{M}g_{ab}+2\chi(\pazocal{M}_{ab}-pg_{ab}), \label{mefe1}
\end{equation}
where $G_{ab}= R_{ab} - \frac{1}{2}R g_{ab}$ is the Einstein tensor. When $\chi$ vanishes, one regains the unmodified form of Einstein's field equations.

\section{\label{sec3} Einstein-Maxwell system in $f(R,T)$ gravity}

We consider a static charged sphere filled with a perfect fluid in a spherically symmetric static spacetime metric
\begin{equation}
ds_{-}^2 = -e^{2\nu(r)} dt^2 + e^{2\mu(r)}dr^2 + r^2(d\theta^2 + \sin^2\theta d\phi^2),\label{intm1}
\end{equation}
in standard coordinates $x^i = (t, r, \theta, \phi)$. The undetermined functions $\nu(r)$ and $\mu(r)$ can be obtained by solving (\ref{mefe1}) together with Maxwell's equations
\begin{equation}
\hspace{1.5cm}F_{[ab,c]} = 0, ~~~~~\left[e^{-(\nu+\mu)}r^2E\right]' =  {4\pi}\sigma e^{\mu} r^2.\label{emax}
\end{equation}
Spherical symmetry implies that $F_{tr}$ is the only non-vanishing component of the electromagnetic field tensor. Using Eq.~(\ref{emax}), we write the electric field intensity as  
\begin{equation}
\hspace{2cm}E = \frac{e^{(\nu+\mu)}}{r^2}q(r),\label{ef}
\end{equation}
where the total charge $q(r)$ contained within the sphere of radius $r$ is defined as
\begin{equation}
\hspace{2cm}q(r) = 4\pi \int_0^r\sigma r^2 e^\mu dr.\label{chareq}
\end{equation}

In the natural unit system having $G = c = 1$, using Eq.~(\ref{chareq}), the Einstein-Maxwell field Eqs.~(\ref{mefe1}) and (\ref{emax}) yield
\begin{eqnarray}
{8\pi}\left(\rho +\frac{q^2}{8\pi r^4}\right) + \chi(3\rho-p)\nonumber\\
= -\frac{1}{r^2}+\frac{1}{r^2}\frac{d}{dr}\left(re^{-2\mu}\right),\label{e1} \\
{8\pi}\left(p-\frac{q^2}{8\pi r^4}\right) + \chi(-\rho+3p) \nonumber\\ 
= e^{-2\mu} \Bigl(\frac{2\nu'}{r}+\frac{1}{r^2}\Bigr) -\frac{1}{r^2},\label{e2}\\
{8\pi}\left(p+\frac{q^2}{8\pi r^4}\right) + \chi(-\rho+3p) \nonumber\\
= e^{-2\mu} \Bigl(\nu''+\nu'^2-\nu'\mu' +\frac{\nu'}{r}-\frac{\mu'}{r}\Bigr)\label{e3},\\
4\pi\sigma =\frac{e^{-\mu}}{r^2}\frac{dq}{dr}.\label{e4}
\end{eqnarray}
The prime ($'$) denotes differentiation with respect to the radial parameter $r$. Subtracting Eq.~(\ref{e3}) from Eq.~(\ref{e2}), we obtain
\begin{equation}
\frac{2q^2}{r^4} = e^{-2\mu} \left(\nu''+\nu'^2-\nu'\mu'-\frac{\nu'}{r}-\frac{\mu'}{r}-\frac{1-e^{2\mu}}{r^2}\right). \label{charge}
\end{equation}
It is noted that all the physical quantities like matter density, pressure, and charge density can be evaluated by solving the system of equations (\ref{e1})-(\ref{charge}). To solve the system, the metric potential $\mu(r)$ is assumed in the most general form of the Buchdahl-VT \cite{Buchdahl19591027,Vaidya1982325} ansatz
\begin{equation}
\hspace{1.5cm}e^{\mu} = \sqrt{\frac{1+ f(r)}{1 - \frac{r^2}{C^2}}}\label{lm2},
\end{equation}
where $C$ is an arbitrary constant. Setting $f(r)=0$ in (\ref{lm2}), it is possible to obtain the Schwarzschid interior solution for an incompressible fluid sphere, as will be shown later.

Eq.~(\ref{charge}) is a second order differential equation. To obtain a tractable form, at this stage, we make a coordinate transformation $x^2=1-\frac{r^2}{C^2}$ and introduce a new variable as in ref.~\cite{Sharma202179}
\begin{equation}
\hspace{1.5cm}e^\nu (1+f)^{-\frac{1}{4}} = \psi(x),\label{psi}
\end{equation}
so that Eq.~(\ref{charge}) takes the form
\begin{eqnarray}
&& \frac{d^2\psi}{dx^2}+\left[\frac{f_{xx}}{4(1+f)}-\frac{5f^2_x}{16(1+f)^2}+\frac{xf_x}{2(1-x^2)(1+f)}\right.\nonumber\\
&& \left.+\frac{f}{(1-x^2)^2}-\frac{2q^2(1+f)}{C^2(1-x^2)^3}\right]\psi = 0,\label{trans}
\end{eqnarray}
where $f_x$ represents the first order derivative with respect to $x$.  
The charged analogue of the Schwarzschild solution demands the linearity of $\psi(x)$, i.e. $\frac{d^2\psi}{dx^2}$ must vanish \cite{Sharma202179} which implies 
\begin{eqnarray}
q^2(x) &=& \frac{C^2(1-x^2)^3f_{xx}}{8(1+f)^2}-\frac{5C^2(1-x^2)^3f^2_x}{32(1+f)^3}+\frac{C^2x(1-x^2)^2f_x}{4(1+f)^2}\nonumber\\
&&+\frac{C^2(1-x^2)f}{2(1+f)},\label{Elec}
\end{eqnarray}
and hence
\begin{equation}
\hspace{2cm}\psi(x) = a -b x, \label{psisol}
\end{equation}
where $a$ and $b$ are integration constants. 
The expression (\ref{Elec}) in terms of the radial parameter $r$ takes the form
\begin{eqnarray}
q^2(r) &=& \frac{r^4(C^2-r^2)f''}{8C^2(1+f)^2}-\frac{5r^4(C^2-r^2)f'^2}{32C^2(1+f)^3}-\frac{r^3(3C^2-2r^2)f'}{8C^2(1+f)^2}\nonumber\\
&&+\frac{r^2f}{2(1+f)},\label{efi1}
\end{eqnarray}
which ensures that $q(r)$ is well behaved at $r=0$ as well as at all interior points of the star for any particular choice of $f(r)$.
Consequently, the spacetime metric of a  static and spherically symmetric object in the presence of an electric field is obtained as
\begin{eqnarray}
ds_{-}^2 &=& -(1+ f(r))^\frac{1}{2}\left(a -b\sqrt{1- \frac{r^2}{C^2}}\right)^2 dt^2 + \frac{1+ f(r)}{1 - \frac{r^2}{C^2}} dr^2 \nonumber\\
&&+ r^2(d\theta^2 + \sin^2\theta d\phi^2).\label{intmcomplt}
\end{eqnarray}
The constants ($a$, $b$ and $C$) can be determined by matching this solution to the exterior Reissner-Nordstr\"om metric at the boundary. A physically viable $f(R,T)$ gravity model can be obtained by choosing $f(r)$ suitably. 

\subsection{\label{subsection:1}Uncharged case: ($q=0$ which implies $f(r)=0$)}

For $f(r)=0$, Eq.~(\ref{Elec}) shows that the electric field vanishes. Using Eq.~(\ref{e1}) and (\ref{e2}), the density and pressure in this case are obtained as
\begin{eqnarray}
\rho &=& \frac{3(4\pi +\chi)\sqrt{1-r^2/C^2}-8a(3\pi+ \chi)}{4C^2(\sqrt{1-r^2/C^2}-2a)(2\pi +\chi)(4\pi +\chi)},\label{Schden}\\
p &=& \frac{3\chi \sqrt{1-r^2/C^2}-8\pi(-a+3b\sqrt{1-r^2/C^2})}{8C^2\left(b\sqrt{1-r^2/C^2}-a\right)(2\pi +\chi)(4\pi +\chi)}.\label{Schpres}
\end{eqnarray}
Obviously, for $\chi = 0$, Eq.~(\ref{Schden}) and (\ref{Schpres}) represent the density and pressure in Einstein's gravity for an uncharged incompressible fluid sphere. At $r=0$, Eq.~(\ref{Schpres}) can be rewritten in the form 
\begin{equation}
\hspace{2cm}p_c = \frac{(p_{c})_{0} - \frac{3\chi}{8\pi C^2 (a-b)}}{(2 +\chi/\pi)(4\pi+\chi)}, \label{cp}
\end{equation}
where $(p_{c})_{0} = \frac{3b-a}{(a-b)C^2}$ represents the central pressure in Einstein's gravity ($\chi =0$). 

From Eq.~(\ref{cp}), we note the following:\\
(i) The central pressure diverges if 
\begin{inparaenum}[(1)]
\item  $a = b$  or, ~
\item  $\chi = -2\pi$ or, $-4\pi$;
\end{inparaenum} \\
(ii) The central pressure vanishes if $\chi = \frac{8\pi}{3}(3b-a)$ which implies that $\chi$ is positive or negative depending on whether $b > a/3$ or $b < a/3$.\\
(iii) Depending on whether $\chi$ is positive or negative, either $p_{c} > (p_{c})_{0}$ or $p_{c} < (p_{c})_{0}$, i.e. the respective values of the central pressure is greater or less in $f(R,T)$ gravity than Einstein's gravity.\\
(iv) The bound on the dimensionless parameter $\chi$ is obtained in the form 
\begin{equation}
\hspace{2cm}-4\pi ~< \chi ~< ~\frac{8\pi}{3}(3b-a). \label{chirange}
\end{equation}

Now the $p(r=R) = 0$) condition determines the constant $R=C\sqrt{1-\frac{a^2}{9b^2}}$ in Einstein's gravity which is the radius of the star. For positive pressure, we must also have $b > a/3$. In $f(R,T)$ gravity, the vanishing of pressure condition yields  
\begin{equation}
\hspace{2cm}b = \frac{1}{2}\left(\frac{\chi}{4\pi}+1\right), \label{const}
\end{equation}
which leads us to an interesting conclusion that $\chi$ must vanish for Schwarzschild's interior incompressible solution for which the constant takes the value $b=1/2$. This issue will be further taken up in Sec.~(\ref{sec5}). 

Consequently, using Eq.~(\ref{intmcomplt}), the interior line element for an uncharged compact object in $f(R,T)$ gravity can be written  as
\begin{eqnarray}
ds_{-}^2 &=& -\left[\frac{3}{2}\sqrt{1-\frac{R^2}{C^2}}-\frac{1}{2}\left(\frac{\chi}{4\pi}+1\right)\sqrt{1-\frac{r^2}{C^2}}\right]^2 c^2 dt^2\nonumber\\
&& + \left(1 - \frac{r^2}{C^2}\right)^{-1} dr^2 + r^2(d\theta^2 + \sin^2\theta d\phi^2),\label{intschw1}
\end{eqnarray}
where we have substituted $a=\frac{3}{2}\sqrt{1-\frac{R^2}{C^2}}$. It is easy to note that the usual form of the Schwarzschild interior solution for an incompressible fluid sphere can be regained simply by setting $\chi = 0$.

\subsection{\label{subsection:2}Charged Buchdahl-Vaidya-Tikekar model in $f(R,T)$ gravity: ($f(r) \neq 0$)}

We now consider the Vaidya and Tikekar (VT) \cite{Vaidya1982325} \textit{ansatz} so  that $f(r) = k\frac{r^2}{C^2}$. The VT ansatz is motivated by the observation that the $t=$ constant hypersurface of the associated spacetime, when embedded in a $4$-Euclidean space, turns out to be spheroidal in which the parameter $k$ denotes the departure from the sphericity of associated $3$-space. The $3$-hypersurface becomes flat and spherical for $k = -1, 0$, respectively. The associated spacetime is well behaved for $r < C$ and $k > - 1$. The VT ansatz has been widely used over the years to model compact stars and radiating stellar models, and this geometry is relevant in our construction, particularly in the context of a similar approach adopted by Sharma {\em et al} \cite{Sharma202179}.

With the VT ansatz, using Eq.~(\ref{e1}~-~\ref{e4}), (\ref{lm2}), (\ref{psi}) and (\ref{psisol}), we obtain the metric potentials, energy-density, pressure and charge-density  as
\begin{eqnarray}
e^{\nu(r)} &=& (a-b\sqrt{1-r^2/C^2})(1+kr^2/C^2)^{1/4}, \label{mp1} \\
e^{\mu(r)} &=& \sqrt{\frac{1+\frac{kr^2}{C^2}}{1-\frac{r^2}{C^2}}}, \label{mp2} \\
\rho &=& \frac{1}{D(r,k,\chi)} \times \Bigl[\left\{b(C^2-r^2)-aC^2\sqrt{1-r^2/C^2}\right\} \Bigr. \nonumber\\
&& \Bigl.\left\{3C^2kr^2\left[4\pi(10+11k)+3\chi(2+5k)\right]+ 12C^4 \right.\Bigr. \nonumber\\ 
&& \Bigl.\left.\left[8\pi(1+k)+\chi(2+3k)\right]+k^2 r^4 \bigl[4\pi(1+4k)- \Bigr.\bigr.\right. \nonumber\\
&& \Bigl.\bigl.\left. \chi(11-4k)\bigr]\right\} -8aC^2\chi \sqrt{1-r^2/C^2} \Bigr.\nonumber\\
&& \Bigl.(C^2+kr^2)^2\Bigr], \label{fe1}\\
p &=& \frac{1}{2D(r,k,\chi)} \times \Bigl[\left\{b(C^2-r^2)-aC^2\sqrt{1-r^2/C^2}\right\} \Bigr. \nonumber\\ 
&& \Bigl. \left\{6C^2kr^2\left[\chi(k-22)-12\pi(k+6)\right]+24C^4 \Bigr.\right. \nonumber\\
&& \Bigl.\left.\left[\chi(k-2)-8\pi\right]-k^2 r^4\bigl[8\pi(4k+25)+2\chi \Bigr.\bigr.\right. \nonumber\\
&& \Bigl.\bigl.\left.(4k+37)\bigr] \right\}-16aC^2\sqrt{1-r^2/C^2}(8\pi+3\chi) \Bigr.\nonumber\\
&& \Bigl.(C^2+kr^2)^2\Bigr], \label{fe2} \\
\sigma &=& \frac{\sqrt{k(C^2-r^2)}}{8\pi\sqrt{2}(C^2+k r^2)^3\sqrt{C^2(2-k)+k(7+4 k)r^2}} \times \nonumber\\
&& \Bigl[3 C^4(2-k)+(4 C^2+k r^2)(7+4 k)k r^2\Bigr], \label{fe3}
\end{eqnarray}
where
\begin{eqnarray}
D(r,k,\chi)&=& 32(C^2+kr^2)^3(2\pi+\chi)(4\pi+\chi)\bigl[b(C^2-r^2)\bigr.\nonumber\\
&&\bigl. -aC^2\sqrt{1-r^2/C^2}\bigr].\nonumber
\end{eqnarray}

At the centre $r=0$, Eq.~(\ref{fe1}~-~\ref{fe3}) take the form
\begin{eqnarray}
\rho_c &=& \frac{1}{8(a-b)C^2(2\pi+\chi)(4\pi+\chi)} \times \Bigl[24\pi(a-b)(1+k) \Bigr.\nonumber\\
&& \Bigl. +\chi\left\{a(8+9k)-3b(2+3k)\right\}\Bigr],\label{mcd}\\
p_c &=& \frac{8\pi(3b - a)+3\chi\left\{2b+k(a-b)\right\}}{8(a-b)C^2(2\pi+\chi)(4\pi+\chi)}\label{mcp},\\
\sigma_c &=& \frac{3}{8\pi C^2}\sqrt{\frac{k(2-k)}{2}}.\label{mccd}
\end{eqnarray}
Obviously the central density ($\rho_c$) will be positive if  
\begin{equation}
\hspace{2cm}k > \frac{2\left[3(b-a)(4\pi+\chi)-a\chi\right]}{3(a-b)(8\pi+3\chi)}.\nonumber
\end{equation}
Moreover the positive central pressure ($p_c$) for the above bound on $k$ implies 
\begin{equation}
\hspace{2cm}\frac{(3b-a)}{8C^2(a-b)} > 0. 
\end{equation}
The above bound further suggests that $a > b > a/3$. For a well behaved stellar configuration, $k$ can take values within the range $-1 < k <\infty$.

Using Eq.~(\ref{efi1}), the charge contained within a radial distance $r$ is obtained as
\begin{equation}
\hspace{2cm}q^2(r) =  \frac{kr^6\left[C^2(2-k) + k(7 + 4k)r^2\right]}{8(C^2 + kr^2)^3},\label{efil2}
\end{equation}
which clearly vanishes at the centre. It is also interesting to note that the charge is zero for $k=0$.

\section{Boundary conditions and fixation of the constants}
\label{sec4}
To evaluate the physical quantities, we need to fix the constants ($a,b,C$) for given values of $\chi$ and $k$. The constants can be determined by utilizing the appropriate boundary conditions, as discussed below.

The exterior spacetime of the static charged object is described by the Reissner-Nordstr\"om metric
\begin{eqnarray}
ds_{+}^2 &=& -\left(1-\frac{2M}{r}+\frac{Q^2}{r^2}\right)dt^2 +\left(1-\frac{2M}{r}+\frac{Q^2}{r^2}\right)^{-1} dr^2\nonumber\\
&&+ r^2(d \theta^2 + \sin^2 \theta d\phi^2), \label{Vm}
\end{eqnarray}
where $M$ and $Q$ represent the total mass and charge, respectively. The matching conditions at the boundary $r=R$ are the continuity of the metric potentials $e^\nu,~ e^\mu$ and the vanishing of pressure at the boundary i.e., $p(r=R) = 0$. Noting that $m(R)=M$ and $q(R)=Q$, these conditions imply
\begin{eqnarray}
1-2u +\alpha^2 u^2 &=& (1+ k n)^\frac{1}{2}(a -b\sqrt{1- n})^2,\label{bc1}\\
1-2u+\alpha^2 u^2 &=& \frac{1 - n}{1+ k n},\label{bc2}\\
16\left(1+nk\right)^2 &=& \Bigl[\frac{b(1-n)-a\sqrt{1-n}}{8\pi b(1-n)+3\chi a\sqrt{1-n}}\Bigr] \times \nonumber\\ 
&& \Bigl[\chi\bigl\{24(k-2)+6nk(k-22) -2n^2k^2 \Bigr.\bigr.\nonumber\\
&& \Bigl.\bigl. (37+4k)\bigr\} -8\pi\bigl\{8+nk(22+9k) \Bigr. \bigr.   \nonumber\\
&& \Bigl. \bigl. +n^2k^2(9+4k)\bigr\}\Bigr],\label{bc3}
\end{eqnarray}
where we have substituted $n=\frac{R^2}{C^2}$, $u=\frac{M}{R}$ and $\alpha^2 = \frac{Q^2}{M^2}$ for simplicity. Using Eq.~(\ref{bc2}), we determine the constant
\begin{equation}
\hspace{2cm}C = R\sqrt{\frac{1+ky}{1-y}},\label{const1}
\end{equation}
where $y = 1-2u +\alpha^2 u^2$.

Using Eq.~(\ref{bc1}) and (\ref{bc3}), we evaluate the remaining two constants as
\begin{eqnarray}
b &=& \frac{y}{8\sqrt{1-n}(8\pi+3\chi)(1+kn)^2\sqrt{y\sqrt{1+kn}}} \times \nonumber\\
&& \Bigl[(4\pi+\chi) \left\{n^2k^2(4k+9)+nk(9k+22)+8\right\} \Bigr. \nonumber\\
&& \Bigl. +4\chi(1+nk)(nk-3k-2)\Bigr], \label{const2} \\
a &=& b\sqrt{1-n}+\frac{y}{\sqrt{y\sqrt{1+kn}}}.\label{const3}
\end{eqnarray}
Note that all constants thus are given in terms of $k$, $M$, $R$, $Q$ and $\chi$. It is interesting to note that in the uncharged case with $k=0$, the above constants take the form
\begin{eqnarray}
a &=& \frac{3(4\pi+\chi)\sqrt{1-2u}}{(8\pi+3\chi)}, \label{const4}\\
b &=& \frac{4\pi}{(8\pi+3\chi)}, \label{const5} \\
C &=& R \sqrt{\frac{1}{2u}}, \label{const6}
\end{eqnarray}
which is exactly the same as in Sec.\ref{subsection:1} if one sets $\chi=0$.

By setting $q(R) = Q$, we rewrite Eq.~(\ref{efil2}) in the form
\begin{equation}
\hspace{2cm}\alpha^2 u^2 n =\frac{kn^3[2-k+(7+4k)kn]}{8(1+kn)^3},\label{ch1}
\end{equation}
and substitute the value of $n$ to obtain
\begin{equation}
\hspace{1cm}k = \frac{8\alpha^{2}}{\left(2-u\alpha^{2}\right)\left[\left(2-u\alpha^{2}\right)
+G(u,\alpha^{2})-8\alpha^{2}\right]},\label{keq}
\end{equation}
where $\displaystyle{G(u,\alpha^{2})=\sqrt{\left(2-u\alpha^{2}\right)\left(2+39u\alpha^{2}\right)
-24\alpha^{2}}}$.
Thus, to have $\alpha =0$, we must have $k=0$ while the converse is also true as can be seen in Eq.~(\ref{ch1}).

\section{\label{sec5}Physical acceptability and analysis of physical quantities}
Any physically acceptable stellar interior solution should have the following features: \\
\begin{inparaenum}[(i)]
\item The density and pressure  should be positive throughout the interior of the star i.e., $\rho,~p > 0$; 
\item the pressure $p$ should vanish at some finite radial distance i.e., $p (r = R) = 0$ and
\item the causality condition should be satisfied throughout the star, which implies that $0 \leq \sqrt{\frac{dp}{d\rho}} \leq 1$.
\end{inparaenum}

To verify whether the above conditions are fulfilled in this model, we consider a hypothetical compact object of a given mass and radius. For this, we take the same set of values as in ref.~\cite{Sharma202179} i.e.,  $M = 1.58~M_\odot$ and  $R = 9.1~$km. Using Eq.~(\ref{const1}) - (\ref{const3}), we evaluate the constants for different values of $k$ and $\chi$. 

\begin{figure}[b]
\includegraphics[width=8cm]{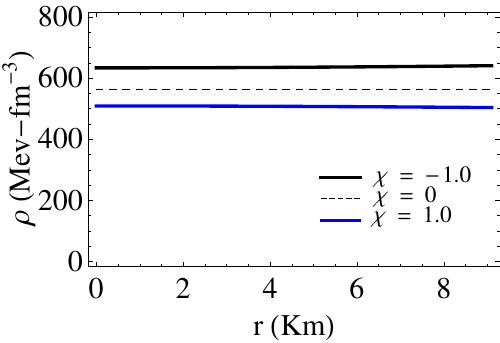}
\caption{\label{fig1} Radial variation of energy density $\rho$ with $\chi$ for an uncharged compact star.}
\end{figure}

\begin{figure}[t]
\includegraphics[width=8cm]{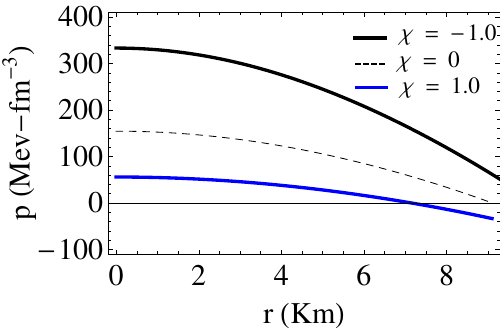}
\caption{\label{fig2} Radial variation of pressure $p$ with $\chi$ for an uncharged compact star. Interestingly, a physically meaningful model is obviously possible only for $\chi = 0$.}
\end{figure}

Let us first consider the feasibility of an uncharged compact object in $f(R,T)$ gravity. Making use of the data given in Table~[\ref{tab1}], we plot the radial variation of matter density and pressure, which are shown in Figs.~(\ref{fig1}) and (\ref{fig2}). Clearly, in Fig.~(\ref{fig2}), we note that the model becomes unphysical for $\chi \neq 0$. In other words, Schwarzschild's interior solution for an incompressible fluid provides an extreme case which cannot be further modified. 

\begin{table}[t]
\caption{\label{tab1}
Values of the model parameters for an uncharged ($k = 0$) compact star of mass $M = 1.58~M_\odot$ and  radius $R = 9.1~$km.}
\begin{center}  
\begin{tabular}{cccc}
\hline
$C$ & $\chi$ & $a$ & $ b$ \\
\hline
& -1.0   & 1.0950  &  0.5678 \\
\hhline{~---}
\multirow{1}{*}{12.7152}  & 0  &  1.0476   &  0.5 \\
\hhline{~---}
& 1.0    & 1.0104   & 0.4467  \\
\hline
\end{tabular}
\end{center}
\end{table}     

\begin{figure}
\begin{minipage}{0.45\columnwidth}
\centering
\includegraphics[width=\textwidth]{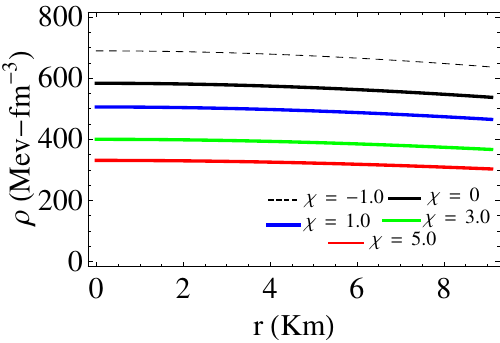}
\caption{Energy density $\rho$ plotted against the radial distance $r$ with $k = 0.1$.}
\label{f1}
\end{minipage}
\hspace{0.5cm}
\begin{minipage}{0.45\columnwidth}
\centering
\includegraphics[width=\textwidth]{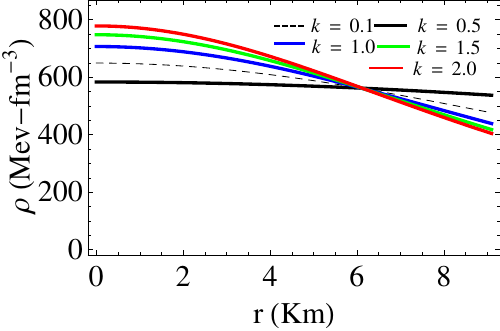}
\caption{Energy density $\rho$ plotted against the radial distance $r$ with $\chi = 0$.}
\label{f2}
\end{minipage}
\end{figure}

\begin{figure}
\begin{minipage}{0.45\columnwidth}
\centering
\includegraphics[width=\textwidth]{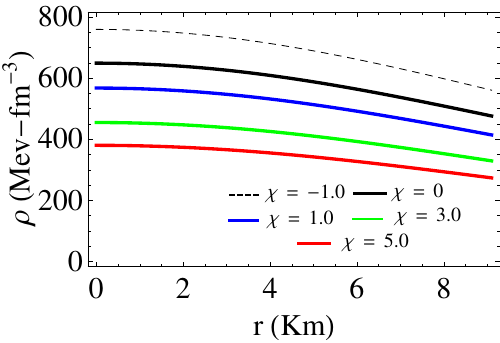}
\caption{Energy density $\rho$ plotted against the radial distance $r$ with $k = 0.5$.}
\label{f3}
\end{minipage}
\hspace{0.5cm}
\begin{minipage}{0.45\columnwidth}
\centering
\includegraphics[width=\textwidth]{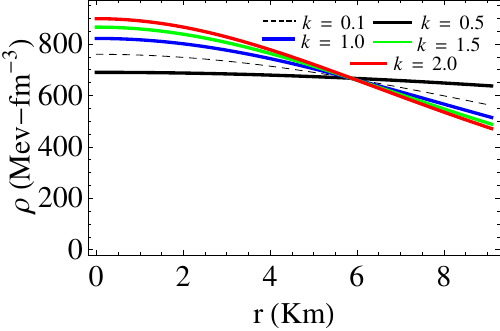}
\caption{Energy density $\rho$ plotted against the radial distance $r$ with $\chi = -1.0$.}
\label{f4}
\end{minipage}
\end{figure}

\begin{figure}
\begin{minipage}{0.45\columnwidth}
\centering
\includegraphics[width=\textwidth]{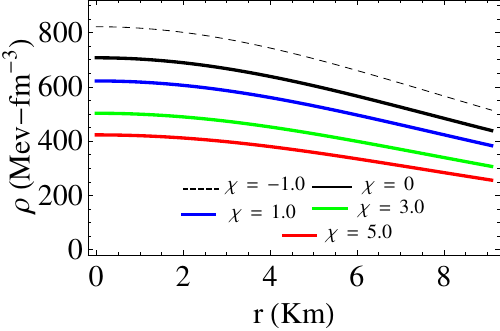}
\caption{Energy density $\rho$ plotted against the radial distance $r$ with $k = 1.0$.}
\label{f5}
\end{minipage}
\hspace{0.5cm}
\begin{minipage}{0.45\columnwidth}
\centering
\includegraphics[width=\textwidth]{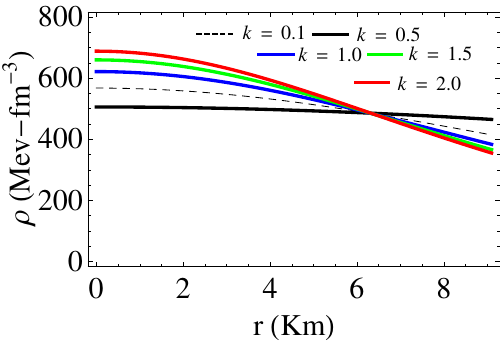}
\caption{Energy density $\rho$ plotted against the radial distance $r$ with $\chi = 1.0$.}
\label{f6}
\end{minipage}
\end{figure}

\begin{figure}
\begin{minipage}{0.45\columnwidth}
\centering
\includegraphics[width=\textwidth]{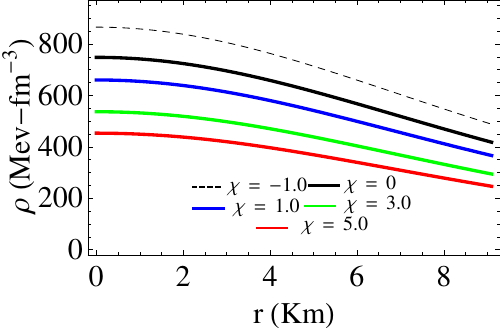}
\caption{Energy density $\rho$ plotted against the radial distance $r$ with $k = 1.5$.}
\label{f7}
\end{minipage}
\hspace{0.5cm}
\begin{minipage}{0.45\columnwidth}
\centering
\includegraphics[width=\textwidth]{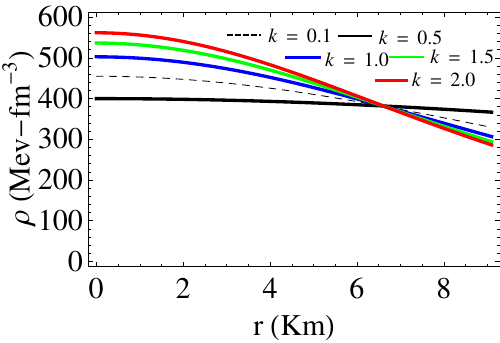}
\caption{Energy density $\rho$ plotted against the radial distance $r$ with $\chi = 3.0$.}
\label{f8}
\end{minipage}
\end{figure}

\begin{figure}
\begin{minipage}{0.45\columnwidth}
\centering
\includegraphics[width=\textwidth]{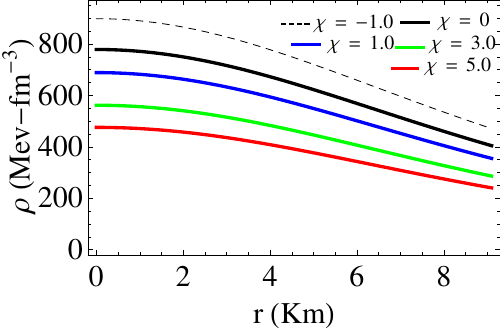}
\caption{Energy density $\rho$ plotted against the radial distance $r$ with $k = 2.0$.}
\label{f9}
\end{minipage}
\hspace{0.5cm}
\begin{minipage}{0.45\columnwidth}
\centering
\includegraphics[width=\textwidth]{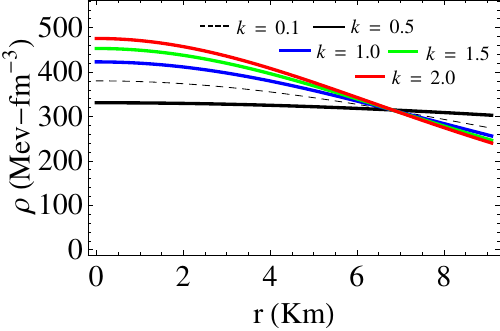}
\caption{Energy density $\rho$ plotted against the radial distance $r$ with $\chi = 5.0$.}
\label{f10}
\end{minipage}
\end{figure}

\begin{figure}
\begin{minipage}{0.45\columnwidth}
\centering
\includegraphics[width=\textwidth]{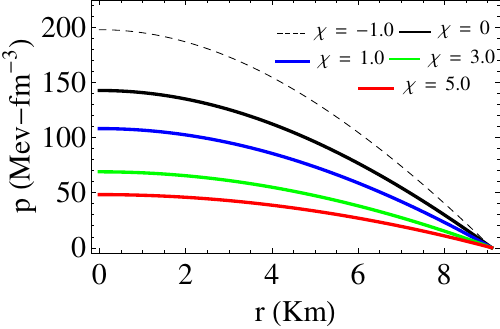}
\caption{Isotropic pressure $p$ plotted against the radial distance $r$ with $k = 0.1$.}
\label{f11}
\end{minipage}
\hspace{0.5cm}
\begin{minipage}{0.45\columnwidth}
\centering
\includegraphics[width=\textwidth]{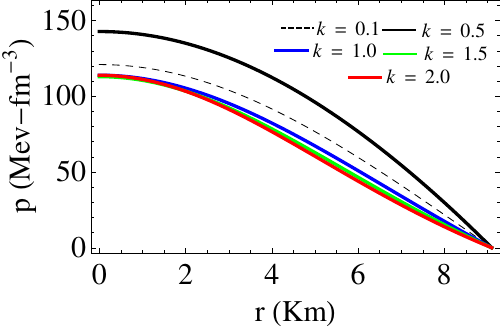}
\caption{Isotropic pressure $p$ plotted against the radial distance $r$ with $\chi = 0$.}
\label{f12}
\end{minipage}
\end{figure}

\begin{figure}
\begin{minipage}{0.45\columnwidth}
\centering
\includegraphics[width=\textwidth]{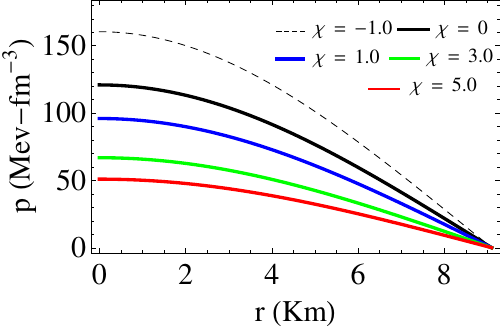}
\caption{Isotropic pressure $p$ plotted against the radial distance $r$ with $k = 0.5$.}
\label{f13}
\end{minipage}
\hspace{0.5cm}
\begin{minipage}{0.45\columnwidth}
\centering
\includegraphics[width=\textwidth]{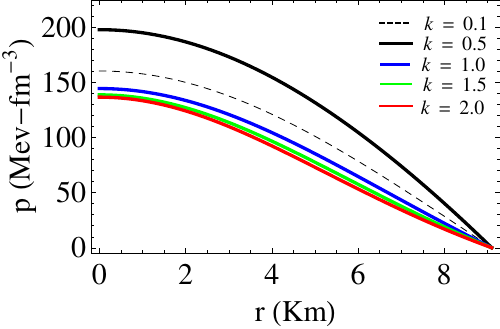}
\caption{Isotropic pressure $p$ plotted against the radial distance $r$ with $\chi = -1.0$.}
\label{f14}
\end{minipage}
\end{figure}

\begin{figure}
\begin{minipage}{0.45\columnwidth}
\centering
\includegraphics[width=\textwidth]{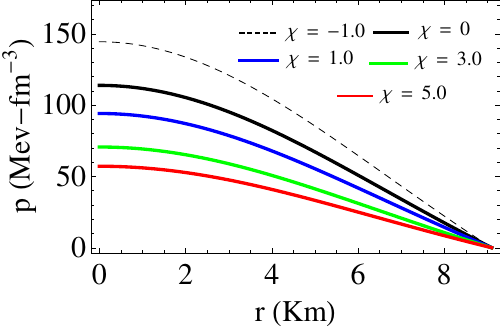}
\caption{Isotropic pressure $p$ plotted against the radial distance $r$ with $k = 1.0$.}
\label{f15}
\end{minipage}
\hspace{0.5cm}
\begin{minipage}{0.45\columnwidth}
\centering
\includegraphics[width=\textwidth]{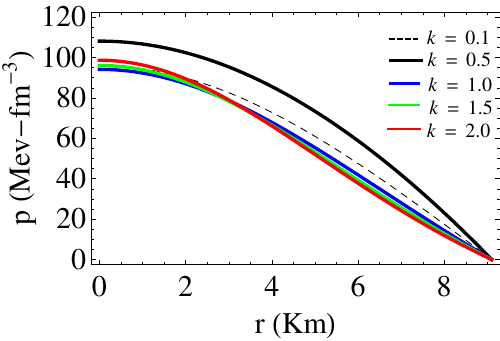}
\caption{Isotropic pressure $p$ plotted against the radial distance $r$ with $\chi = 1.0$.}
\label{f16}
\end{minipage}
\end{figure}

\begin{figure}
\begin{minipage}{0.45\columnwidth}
\centering
\includegraphics[width=\textwidth]{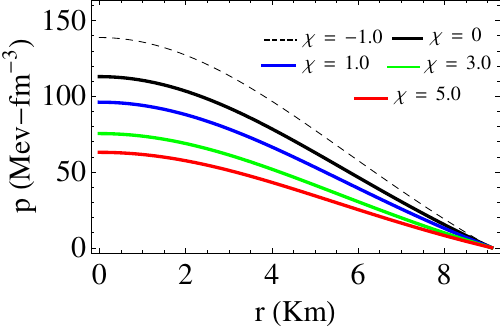}
\caption{Isotropic pressure $p$ plotted against the radial distance $r$ with $k = 1.5$.}
\label{f17}
\end{minipage}
\hspace{0.5cm}
\begin{minipage}{0.45\columnwidth}
\centering
\includegraphics[width=\textwidth]{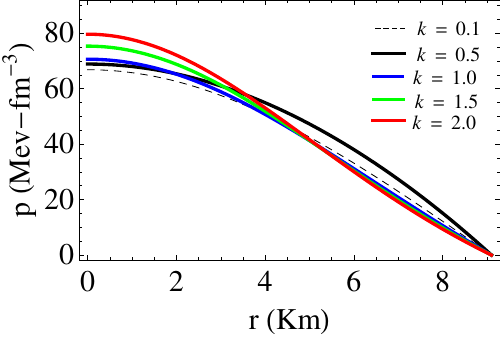}
\caption{Isotropic pressure $p$ plotted against the radial distance $r$ with $\chi = 3.0$.}
\label{f18}
\end{minipage}
\end{figure}

\begin{figure}
\begin{minipage}{0.45\columnwidth}
\centering
\includegraphics[width=\textwidth]{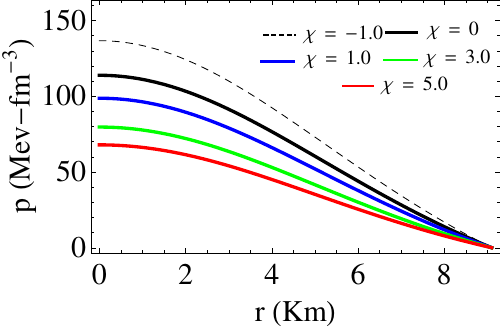}
\caption{Isotropic pressure $p$ plotted against the radial distance $r$ with $k = 2.0$.}
\label{f19}
\end{minipage}
\hspace{0.5cm}
\begin{minipage}{0.45\columnwidth}
\centering
\includegraphics[width=\textwidth]{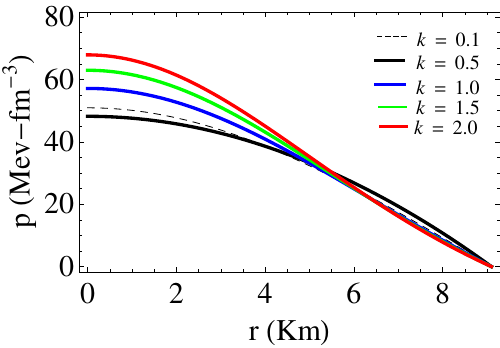}
\caption{Isotropic pressure $p$ plotted against the radial distance $r$ with $\chi = 5.0$.}
\label{f20}
\end{minipage}
\end{figure}

Let us now consider the charged case. For a charged ($k$ or $q(r) \neq 0$) compact object with the mass and radius given above, the values of the constants are given in Table~[\ref{tab2}]. In the table, we note that all values of $a$ are less than $3/2$, and values of $b$ are greater than $a/3$ as well as greater than $1/2$ [as discussed in Sec.\ref{subsection:1}]. With this set of values, we investigate the behaviour of the matter density and pressure for different choices of $k$ and $\chi$ as shown in Fig.~\ref{f1}-\ref{f20}. The plots indicate that the model parameters are regular and well behaved at all interior points of the compact object. Some features of the model are discussed below:\\
\begin{enumerate}[(i)]
\item Fig.~\ref{f1}, \ref{f3}, \ref{f5}, \ref{f7} and \ref{f9} show that the radial variation of matter density monotonically decreases as one goes outward. 
\item Fig.~\ref{f2}, \ref{f4}, \ref{f6}, \ref{f8} and \ref{f10} show that as $\chi$ increases the matter density gradually decreases near the core or the central region. Irrespective of the signature and values of $\chi$, all the density curves cross the constant density line at $r\sim 6~$ km, which imply that any uniform density stellar configuration has the largest value at the boundary.
\item Fig.~\ref{f11}, \ref{f13}, \ref{f15}, \ref{f17} and \ref{f19} show that as the charge increases, the central pressure decreases monotonically. 
\item Fig.~\ref{f12}, \ref{f14}, \ref{f16}, \ref{f18} and \ref{f20} show the radial variation of the isotropic pressure for different values of $\chi$. Fig.~\ref{f12} indicates that the negative values of $\chi~$ exert more pressure at the core region than positive values. Interestingly, $k$ and negative values of $\chi$ seem to have opposite effects.  
\end{enumerate}

\begin{table*}
\caption{\label{tab2}
Values of the model parameters for different choices of $k$ and $\chi$ for a hypothetical charged compact star of mass $M = 1.58~M_\odot$ and radius $R = 9.1~$km.}
\begin{center}  
\begin{tabular}{||cccc||cccc||cccc||}
\hline
\multicolumn{12}{||c||}{$\chi = -1.0$} \\
\hline
$k$ & $C$ & $a$ & $ b$ & $k$ & $C$ & $a$ & $ b$ & $k$ & $C$ & $a$ & $ b$  \\
\hline
0.1 & 13.0983 & 1.1047 & 0.5706 & 1.0 & 16.0447 & 1.2042 & 0.6590  &\multirow{2}{*}{2.0} &\multirow{2}{*}{18.7246} &\multirow{2}{*}{1.3340}  &\multirow{2}{*}{0.7828} \\
0.5 & 14.4983 & 1.1460 & 0.5982 & 1.5 & 17.4398 & 1.2676 & 0.7124 &  &   &   & \\
\hline
\multicolumn{12}{||c||}{$\chi = 0$} \\
\hline
$k$ & $C$ & $a$ & $ b$ & $k$ & $C$ & $a$ & $ b$ & $k$ & $C$ & $a$ & $ b$  \\
\hline
0.1 & 13.0983 & 1.0547 & 0.5011 & 1.0 & 16.0447 & 1.1299 & 0.5607  &\multirow{2}{*}{2.0} &\multirow{2}{*}{18.7246} &\multirow{2}{*}{1.2323}  &\multirow{2}{*}{0.6619} \\
0.5 & 14.4983 & 1.0853 & 0.5202 & 1.5 & 17.4398 & 1.1796 & 0.6093 &  &   &   & \\
\hline
\multicolumn{12}{||c||}{$\chi = 1.0$} \\
\hline
$k$ & $C$ & $a$ & $ b$ & $k$ & $C$ & $a$ & $ b$ & $k$ & $C$ & $a$ & $ b$  \\
\hline
0.1 & 13.0983 & 1.0154 & 0.4464 & 1.0 & 16.0447 & 1.0715 & 0.4897  &\multirow{2}{*}{2.0} &\multirow{2}{*}{18.7246} &\multirow{2}{*}{1.1524}  &\multirow{2}{*}{0.5704} \\
0.5 & 14.4983 & 1.0375 & 0.4589 & 1.5 & 17.4398 & 1.1104 & 0.5282 &  &   &   & \\
\hline
\multicolumn{12}{||c||}{$\chi = 3.0$} \\
\hline
$k$ & $C$ & $a$ & $ b$ & $k$ & $C$ & $a$ & $ b$ & $k$ & $C$ & $a$ & $ b$  \\
\hline
0.1 & 13.0983 & 0.9575 & 0.3659 & 1.0 & 16.0447 & 0.9854 & 0.3852  &\multirow{2}{*}{2.0} &\multirow{2}{*}{18.7246} &\multirow{2}{*}{1.0346}  &\multirow{2}{*}{0.4356} \\
0.5 & 14.4983 & 0.9672 & 0.3685 & 1.5 & 17.4398 & 1.0085 & 0.4087 &  &   &   & \\
\hline
\multicolumn{12}{||c||}{$\chi = 5.0$} \\
\hline
$k$ & $C$ & $a$ & $ b$ & $k$ & $C$ & $a$ & $ b$ & $k$ & $C$ & $a$ & $ b$  \\
\hline
0.1 & 13.0983 & 0.9169 & 0.3095 & 1.0 & 16.0447 & 0.9251 & 0.3120  &\multirow{2}{*}{2.0} &\multirow{2}{*}{18.7246} &\multirow{2}{*}{0.9520}  &\multirow{2}{*}{0.3412} \\
0.5 & 14.4983 & 0.9179 & 0.3052 & 1.5 & 17.4398 & 0.9371 & 0.3250 &  &   &   & \\
\hline
\end{tabular}
\end{center}
\end{table*}

\section{Compactness bound in $f(R,T)$ gravity}
\label{sec6}

The mass to radius ratio, i.e compactness ($M/R$), plays a crucial role in modelling a compact star. The most compact object is a black hole (BH) for which the compactness ratio is $\frac{1}{2}$. For non-BH compact objects, amongst others, the upper bound on $M/R$ in Einstein's gravity was investigated by Sharma \textit{et al} \cite{Sharma202179} in which it has been shown that a charged non-BH could be overcharged as compared to a charged black hole. In the current investigation, we intend to analyze the impact of the modification made in Einstein's gravity on the compactness bound. Recently, in the uncharged case, Pappas \textit{et al} \cite{Pappas2022124014} have extended the Tolman III and VII solutions to $f(R,T)$ gravity and showed that the mass to radius ratio becomes greater than the Buchdahl limit of compactness for positive values of $\chi$. The ratio never exceeds the black hole compactness.

In the charged case, we obtain the bound by demanding that the central pressure must not diverge. From Eq.~(\ref{mcp}), it is evident that $(a - b) \geq 0$ is the condition for the divergence-free central pressure. Substituting the value of $a$, as given in Eq.~(\ref{const3}) in the above requirement, we obtain
\begin{equation}
\hspace{2cm}\frac{y}{\sqrt{y\sqrt{1+kn}}} \geq b\left(1-\sqrt{1-n}\right). \label{cond1}
\end{equation}
For an uncharged ($k = \alpha = 0$) compact object this condition simplifies to 
\begin{equation}
\hspace{2cm}\left(4\pi+\chi\right)\left(3\sqrt{1-2u} - 1\right) + \chi \geq 0 , \label{cond2}
\end{equation}
which, on further simplification, gives  
\begin{equation}
\hspace{2cm}\boxed{
\frac{M}{R} \leq \frac{1}{2} \left[1- \frac{1}{9\left(1+\frac{\chi}{4\pi}\right)^2}\right]}. \label{cond3}
\end{equation}
One can readily retrieve the Buchdahl bound $\frac{M}{R} \leq \frac{4}{9}$ from Eq.(\ref{cond3}) for $\chi = 0$. It should be stressed here that even though $\chi$ has no impact on Schwarzschild's incompressible fluid solution if the constants $a$ and $b$ in Eq.~(\ref{intschw1}) remain arbitrary, Eq.~(\ref{cond3}) provides the compactness bound for the uncharged sphere in $f(R,T)$ gravity.   

\begin{figure}[b]
\includegraphics[width=8cm]{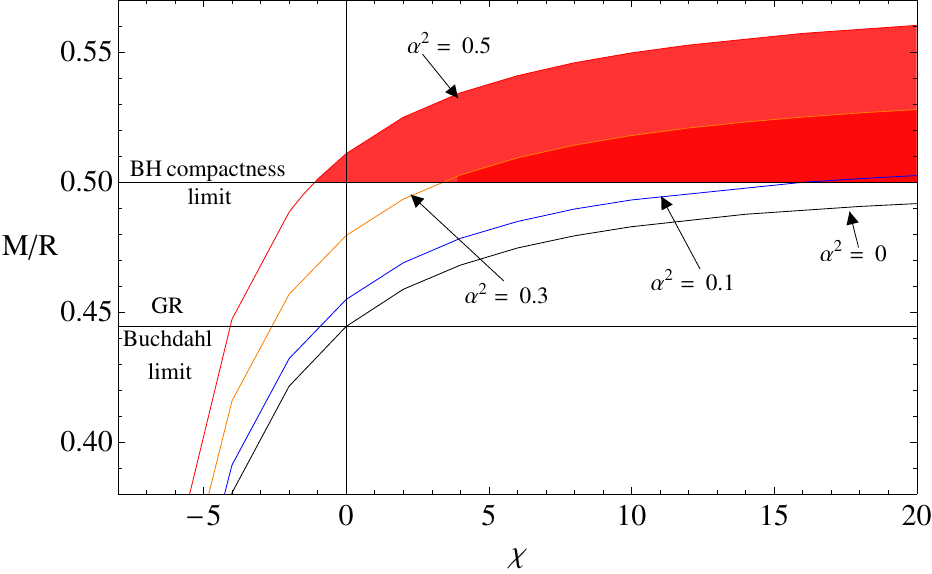}
\caption{\label{f21}Compactness $\frac{M}{R}$ plotted against $\chi$ for different choices of $\alpha^{2}$.}
\end{figure}

For a charged star, condition~(\ref{cond1}) takes the form as
\begin{eqnarray}
\left(4\pi+\chi \right)\Bigl[16\left(1+nk\right)^2\sqrt{1-n}-\left(1-\sqrt{1-n}\right) \Bigr.\nonumber\\
\hspace{0.1cm} \Bigl.\left\{8+nk(9k+22)+n^2k^2(4k+9)\right\}\Bigr]+4\chi\left(1+nk\right) \nonumber\\
\hspace{0cm} \left[k(n-1)(3\sqrt{1-n}-1)+2(k+1)\right] \geq 0. \label{cond4}
\end{eqnarray} 
Eq.~(\ref{bc2}) and (\ref{keq}) suggest that the above condition (\ref{cond4}) is governed by choice of $\alpha$ and $\chi$. Condition~(\ref{cond4}) can be utilized numerically to find a bound on compactness as shown in Fig.~(\ref{f21}). It is noteworthy that while the compactness for an uncharged stellar configuration cannot go beyond black hole compactness, it can go beyond the BH compactness when the star is charged. As the values of $\alpha^{2}$ are increased, the compactness goes beyond the BH compactness limit even for small values of $\chi$.

Having understood the dependency of $\chi$ on compactness numerically, let us now explore the possibility of obtaining an analytic expression yielding similar behaviour, which might be considered as the compactness bound for a charged sphere in $f(R,T)$ gravity analogous to the Buchdahl bound. To achieve this goal, let us assume $k = \kappa \epsilon$ and $\chi = X \epsilon$, where $\left|\epsilon\right| << 1$. This approximation will be valid if the departure from sphericity is small and the modification is moderate. With these assumptions, we have from Eq.~(\ref{keq})
\begin{equation}
\alpha^2 = \frac{2\left[\left(1-k+3ku\right)-\sqrt{\left(1-k+3ku\right)^2 - 5k^2u^2}\right]}{5ku^2} + \pazocal{O}(\epsilon ^2). \label{aeq}
\end{equation} 
Inserting the value of $\alpha ^{2}$ in Eq.~(\ref{cond4}) and retaining terms upto $\pazocal{O}(\epsilon)$, we obtain
\begin{equation}
\hspace{1cm}\boxed{
u = \frac{M}{R} = \frac{\frac{32\pi}{9}\left[\frac{2}{(8\pi - \chi)}-\frac{\chi}{(8\pi-\chi)^2}\right]}{1+\frac{4\pi}{(8\pi-\chi)}\sqrt{\frac{16\alpha ^2}{9}(\frac{3\chi}{8\pi}-2)+(\frac{\chi}{4\pi}-2)^2}}}, \label{cnr}
\end{equation}
which for a charged compact object in Einstein's gravity (i.e. $\chi = 0$)  takes the form
\begin{equation}
\hspace{2cm}u  =  \frac{M}{R}  = \frac{8/9}{1+\sqrt{1-\frac{8\alpha^2}{9}}}. \label{cbl}
\end{equation}
Thus, as far as compactness is concerned, we notice a distinctive behaviour in $f(R,T)$ gravity. Eq.~(\ref{cnr}) may be considered as a charged generalization of the Buchdahl bound in $f(R,T)$ gravity. 

\begin{figure}
\includegraphics[width=8cm]{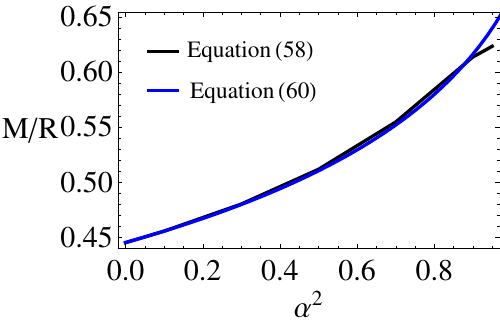}
\caption{\label{f22}Compactness ratio $\frac{M}{R}$ plotted against $\alpha^2$ for $\chi = 0.1$ using Eq.~(\ref{cnr}) and Eq.~(\ref{cond4}).}
\end{figure}

For a small value of $\chi$ (we take $=0.1$), we first evaluate the variation of compactness ($\frac{M}{R}$) with $\alpha^{2}$ by utilizing the condition (\ref{cond4}) which is tabulated in Table~(\ref{tab3}). Fig.~\ref{f22} utilizes the data obtained in Table~(\ref{tab3}) to plot the variation of compactness with $\alpha^2$. The plot is then embedded on a similar plot obtained by utilizing condition (\ref{cnr}). The overlapping of the two plots justifies the validity of our approximation method and the result in Eq.~(\ref{cnr}). 

\begin{table}[t]
\caption{\label{tab3}
Variation of compactness $\frac{M}{R}$ with $\alpha^2$ for $\chi = 0.1$}
\begin{tabular}{cccccccc}
\hline
$\alpha^2$ & 0 & 0.1 & 0.3 & 0.5 & 0.7 & 0.9 & 0.95 \\ 
\hline
$\frac{M}{R}$ & 0.4453 & 0.4557 & 0.4803  & 0.5118 &  0.5547  & 0.6137  & 0.6234 \\ 
\hline 
\end{tabular}
\end{table}

\begin{figure}
\includegraphics[width=8cm]{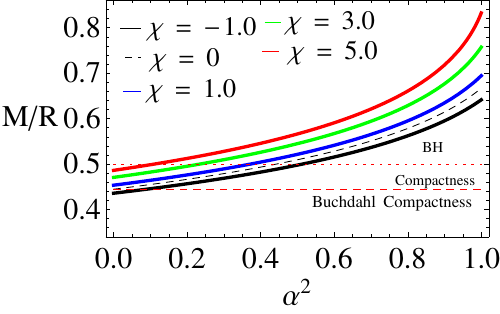}
\caption{\label{f23}Compactness ratio $\frac{M}{R}$ plotted against $\alpha^2$ for different choices of $\chi$.}
\end{figure}

\section{\label{sec7}Concluding remarks}

The Buchdahl bound was studied by Goswami {\em et al} \cite{Goswami2015} leading to new features in stellar objects. We expect to also obtain interesting features in $f(R,T)$ theory for the Buchdahl limit. This paper provides an analysis of the physical behaviour of a charged compact star in $f(R,T)$ gravity. The electromagnetic extension of the Buchdahl bound obtained in $f(R,T)$ gravity is distinct from the results obtained earlier by Sharma {\em et al} \cite{Sharma202179} for general reativity. Our study shows that the compactness can be increased by considering a modification in Einstein's gravity which is further enhanced by the inclusion of charge. While in the absence of charge, the compactness never exceeds the BH compactness $0.5$, even a comparatively small amount of charge together with the impact of the trace of the stress-energy tensor $T$ can exceed the compactness bound beyond the BH limit $0.5$. Whether this is indicative of a more stringent bound on the coupling term $\chi$ demands further probe. A another point to note is that in Eq.~(\ref{cnr}), $u$ will be a positive real quantity if the squared root term in the denominator remains positive. This restricts the charge to mass ratio $\frac{Q^2}{M^2} < \frac{9}{8}\frac{(\chi-8\pi)^2}{4\pi(16\pi-3\chi)}$. Obviously, $\frac{Q^2}{M^2} < 9/8$ for $\chi=0$. To conclude, while, in the absence of charge, the variation of compactness is similar to the results obtained in ref.~\cite{Pappas2022124014}, the presence of charge provides some new insight into the effects $f(R,T)$ gravity on compactness.

\begin{acknowledgements}
RS gratefully acknowledges support from the Inter-University Centre for Astronomy and Astrophysics (IUCAA), Pune, India,  under its Visiting Research Associateship Programme.
\end{acknowledgements}

\section*{Declarations}
\begin{itemize}
\item Funding: Not applicable
\item Conflict of interest/Competing interests: ~The authors declare no conflict of interest.
\item Ethics approval 
\item Consent to participate:~All authors have read and agreed to the published version of the manuscript.
\item Consent for publication:~All authors have read and agreed to the published version of the manuscript.
\item Availability of data and materials: This article's data is accessible within the public domain as specified and duly referenced in the citations.
\item Code availability: Not applicable
\item Authors' contributions: All the authors have contributed equally to the manuscript.
\end{itemize}

\end{document}